\documentclass[submission,copyright,creativecommons]{eptcs}
\providecommand{\event}{ACL2 2022} % Name of the event you are submitting to
\usepackage{underscore}           % Only needed if you use pdflatex.

\usepackage{graphicx}
\usepackage{xcolor} 
\usepackage{amsmath}
\usepackage{amsfonts}
\usepackage{amssymb}
\usepackage{listings}
\definecolor{commentsColor}{rgb}{0.497495, 0.497587, 0.497464}
\definecolor{keywordsColor}{rgb}{0.000000, 0.000000, 0.635294} %{1,0,0}
\definecolor{stringColor}{rgb}{0.558215, 0.000000, 0.135316}
\title{Using ACL2 To Teach Students About Software Testing}
\author{Ruben Gamboa \qquad Alicia Thoney
\institute{University of Wyoming \\ Laramie, Wyoming}
\email{\{ruben,athoney\}@uwyo.edu}
}
\def\titlerunning{Teaching Software Testing with ACL2}
\def\authorrunning{R. Gamboa \& A. Thoney}
\begin{document}
\maketitle

\begin{abstract}
We report on our experience using ACL2 in the classroom to teach students about software testing. The course COSC~2300 at the University of Wyoming is a mostly traditional Discrete Mathematics course, but with a clear focus on computer science applications. For instance, the section on logic and proofs is motivated by the desire to write proofs about computer software. We emphasize that the importance of software correctness falls along a spectrum with casual programs on one end and mission-critical ones on the other. Corresponding to this spectrum is a variety of tools, ranging from unit tests, randomized testing of properties, and even formal proofs. In this paper, we describe one of the major activities, in which students use the ACL2 Sedan's counter-example generation facility to investigate properties of various existing checksum algorithms used in error detection. Students are challenged to state the relevant properties correctly, so that the counter-example generation tool is used effectively in all cases, and ACL2 can find formal proofs automatically in some of those.
\end{abstract}

\section{Introduction}
\label{sec:intro}

At the University of Wyoming, we have been teaching a Discrete Structures course (COSC~2300) every semester since 
the fall of 2016. The course covers all the typical topics from Discrete Mathematics, but with an emphasis on
Computer Science applications to make it more relevant to CS majors, who make up the vast majority of students.
The course is also heavily
influenced by an Honors Course for non-majors developed at the University of Oklahoma by 
Rex Page~\cite{DBLP:journals/corr/abs-1301-5074,DBLP:conf/sigcse/PageG13}. One of those influences is the use of
ACL2 to achieve some of the course learning outcomes.

To be clear, competence in ACL2 is not one of those learning outcomes. Instead, students are expected to master
the usual topics in discrete mathematics: basic proof techniques, induction, inductive data structures (e.g., 
lists and trees), foundational mathematical structures (e.g., sets, relations, functions), modular arithmetic, 
and basic combinatorics. What ACL2 brings to the table is in connecting the abstract concepts of discrete mathematics
with the more practical aspects of software development. 

Again to be clear, we make no attempt to teach software verification in a sophomore-level course. Rather, we
emphasize that there is a spectrum of reliability needs for software. On one end are simple programs that few
people genuinely care about, like simple iPhone games. On the other end are mission- and even life-critical
software, like aerospace, automotive, or health applications. So as software developers, we have a wide range
of tools that can be used to improve the reliability of our software.

At the lower end stand unit test cases, which are widely acknowledged to be a best practice in software development.
Students learn about unit tests in our class, and also in other classes as they progress through their degree.
In ACL2, we expose unit tests through the macro \texttt{check-expect}, a common testing framework that allows the
programmer to provide an expression and its expected return value. Although the user interface is
minimal, this is very much comparable to industrial practice with testing frameworks like JUnit or
Jasmine~\cite{junit5,jasmine}. It is helpful that students have mostly already accepted unit testing as a
necessary evil, so this serves as the context for ACL2.

More confidence in the correctness of the software can be gained by using randomized testing. In this scenario,
the programmer describes some properties about the software, e.g., that the result returned by a particular function 
$f(n)$ is an odd prime whenever $n$ is a positive integer. These properties, of course, are described using 
first-order logic in general, and ACL2 expressions in particular. This serves to suggest one of the applications
of logic to software development: Logic can be used to describe important properties of the software we are writing.
Note that the properties need not form a complete correctness specification. For example, the property above requires 
that $f(n)$ be an odd prime number, but it does not provide enough information to determine which prime is the right 
one. Such properties are sometimes called ``little theories'' about the code, and they can be validated by using
randomized testing. The idea is to test a large number of input values and verify the function satisfies the
property for each one of those inputs; e.g., $f(n_1)$, $f(n_2)$, \dots, $f(n_{1000})$, are all odd prime numbers.
Such properties can be verified in ACL2 using ACL2 testing frameworks, such as \texttt{doublecheck} or 
counter-example generation in ACL2s~\cite{10.1145/1637837.1637844,testing-in-acl2s}.

Even higher levels of confidence in the correctness of the software can be gained by proving that the desired
properties are actually true. Such proofs are notoriously tedious, so they are practical only with a great deal
of automation support. Enter ACL2. It is very helpful that the ACL2 testing frameworks use the same syntax to 
describe properties, whether they are intended for randomized testing or proof. We give students perhaps an
unrealistic expectation when it comes to proof automation, since we work very hard to ensure that all properties
that students are asked to prove are simple enough that ACL2 can prove them automatically. What we want students
to learn is that proving software correctness is possible, and that it is a viable option in industrial applications,
but we do not even attempt to give them the tools necessary to become proficient ACL2 users.

The rest of this paper is organized as follows. In Sect.~\ref{sec:which-acl2} we discuss the pragmatic issue of
choosing an ACL2 distribution for students. Then in Sect.~\ref{sec:how-much-acl2} we discuss the limited subset
of ACL2 that we present to students. Sect.~\ref{sec:acl2-assignment} describes a specific assignment we used that showcases
how ACL2 and property testing can be used to explore a specific set of algorithms, and how the testing frameworks
can demonstrate that certain properties that a programmer may assume are actually not true. Finally, some concluding
remarks are included in Sect.~\ref{sec:conclusions}.

\section{Which ACL2 Distribution?}
\label{sec:which-acl2}

Many, if not most, expert users of ACL2 rely on Emacs and run ACL2 in a terminal emulator, such as an Emacs shell.
However, current students at the University of Wyoming are more accustomed to full-blown IDEs like Visual Studio,
IntelliJ IDEA, or Eclipse. Emacs's typical steep learning curve is not justified for our purposes, so we
have always used more user-friendly distributions of ACL2.

The obvious distribution is ACL2s, the ACL2 Sedan~\cite{DBLP:journals/entcs/DillingerMVM07}. We used this
distribution for a couple of years, and there was much that was right with it. For example, students were not
burdened with worrying about termination of their programs very much, and the counter-example generation feature
made it easier for them to find mistakes in their formal models. Another ACL2s feature that strongly supported
students was the strong typing provided by \texttt{defdata} and \texttt{definec}. Using \texttt{definec} students
can say ``$f$ is a function that maps natural numbers to rationals,'' for example. Users who are accustomed to ACL2's
lack of typing may consider that a feature, but students who grew up on Java and C++ can get quickly lost when
writing a function that processes lists of lists of naturals, for example. On the other hand, the \texttt{defdata}
model for defining types is something that Java and C++ programmers can quickly grasp, and that makes it easier
to succeed in writing programs. It is also an excellent feature that \texttt{defdata} and the counter-example
generation features work so well together, so students learn to program in ACL2 and to test ACL2 programs at the
same time.

However, we decided to move away from ACL2s
based on student feedback. Some students were confused by the fact that the source code was in a different tab
than the program execution, and that at any given time some of the source code was writable but some of it was
immutable. We also encountered mundane problems with installation. The IT department at the university was
flummoxed at the prospect of installing ACL2s for all users, but with a different file space for each user.
This is a complication that ACL2s inherits from Eclipse. And, at least under Windows, there were always problems
with file permissions and folders containing strange characters, such as a space.

Because of this, we started using DrACuLa~\cite{DBLP:conf/acl2/VaillancourtPF06} and the 
DoubleCheck~\cite{10.1145/1637837.1637844} testing framework instead. This worked reasonably well. Students
preferred the user interface of DrACuLa to the sedan's, perhaps because it was designed from the ground up with
students in mind, unlike Eclipse, which is geared to professional programmers. 
Installation was still an issue but not as much as with ACL2s. Again, this is mostly due to Eclipse. In fact,
at least under Windows, DrACuLa uses the same ACL2 executable that is distributed with ACL2s. A minor drawback
of the DoubleCheck framework is worth noting. In ACL2s, randomized testing is performed by inspecting the 
hypothesis of the theorem to derive the possible values of each variable, e.g., that $n$ is a positive integer. 
In DoubleCheck, it is the programmer's responsibility to tell DoubleCheck the universe of values from which to
randomly select values for the variables. Thus, it is possible to write a theorem where $n$ is declared as an
integer, but to test it with only positive values of $n$. The tests may all pass, even if the theorem is not true
when $n=-1$, for example. Students accepted this, but they were never entirely satisfied.

Although installing DrACuLa was easier than installing ACL2s, we eventually moved from DrACuLa to 
ProofPad~\cite{EPTCS114.2}, a web-based interface to ACL2 developed by Caleb Eggensperger, a former student in Rex Page's 
Honors Course at Oklahoma. The advantage of using ProofPad was clear: It is a web-based interface, so it requires
absolutely no installation. On Day 1 of the course, all students are able to run ACL2 from their laptops. The
user interface of ProofPad is clearly influenced by minimalistic design. There are no extra buttons or contextual
menus that could potentially confuse students. ProofPad incorporates DoubleCheck, so it has the same speed bumps
in user experience when it comes to checking versus proving properties.

The fact that ProofPad was entirely available as a web application was its biggest strength, but also a weakness.
The first version of ProofPad used Unix pipes on the web server to communicate between the web server and the underlying ACL2, which 
simply behaved as if it was invoked from the command line. This
was prone to errors, since sometimes ProofPad would miss an ACL2 prompt and the entire application would freeze. 
If ACL2 encountered a hard error, 
for example, ProofPad would sometimes get lost and not
recover. Students complained, naturally enough, about mysterious ``broken socket'' errors. A second release of
ProofPad fixed many of these issues by switching to the ACL2 Bridge for communication~\cite{bridge}. The bridge
is a remarkable and resilient tool that allows other applications to submit events and expressions to ACL2. While
that mostly solved the communication issues, the development of ProofPad stalled and the last release we used had
some UI errors that made it extremely hard to use; e.g., if the input file had a very long line, the ACL2 output
became literally unreadable.

For the past year, we have moved to a solution that is inspired by both ACL2s and ProofPad. Using the ACL2
bridge, we built a back-end that supports ACL2 in Jupyter notebooks, which are accessed through a web browser, 
just like ProofPad\footnote{Perhaps this is the topic of an
upcoming rump session.}. Jupyter is widely used by people in the scientific and data science communities, and
it support numerous back ends to languages such as Python, R, and (now) ACL2. The version of ACL2 that we use
is, in fact, ACL2s. More precisely, it is a vanilla version of ACL2 but with an initialization file that loads
several features originally developed for ACL2s, such as counter-example generation and CCG termination
analysis. Transparently to the students, everything they type is interpreted in the ``ACL2s'' package.

Installation, of course, is a major issue. This solution requires not only a successful installation of ACL2,
but also an installation of a Jupyter server. We assuage student discomfort by providing them with a Docker
image that includes a complete Jupyter server with its own ACL2 server. This means that students need to install
Docker on their laptops, but this has proven to be easier---and it's getting easier, even under Windows, thanks
to the Windows Subsystem for Linux (WSL). The IT department at our university was also able to install Docker on
the lab computers, so this solution is widely available.

\begin{figure}
\begin{center}
\framebox{\includegraphics[scale=0.45]{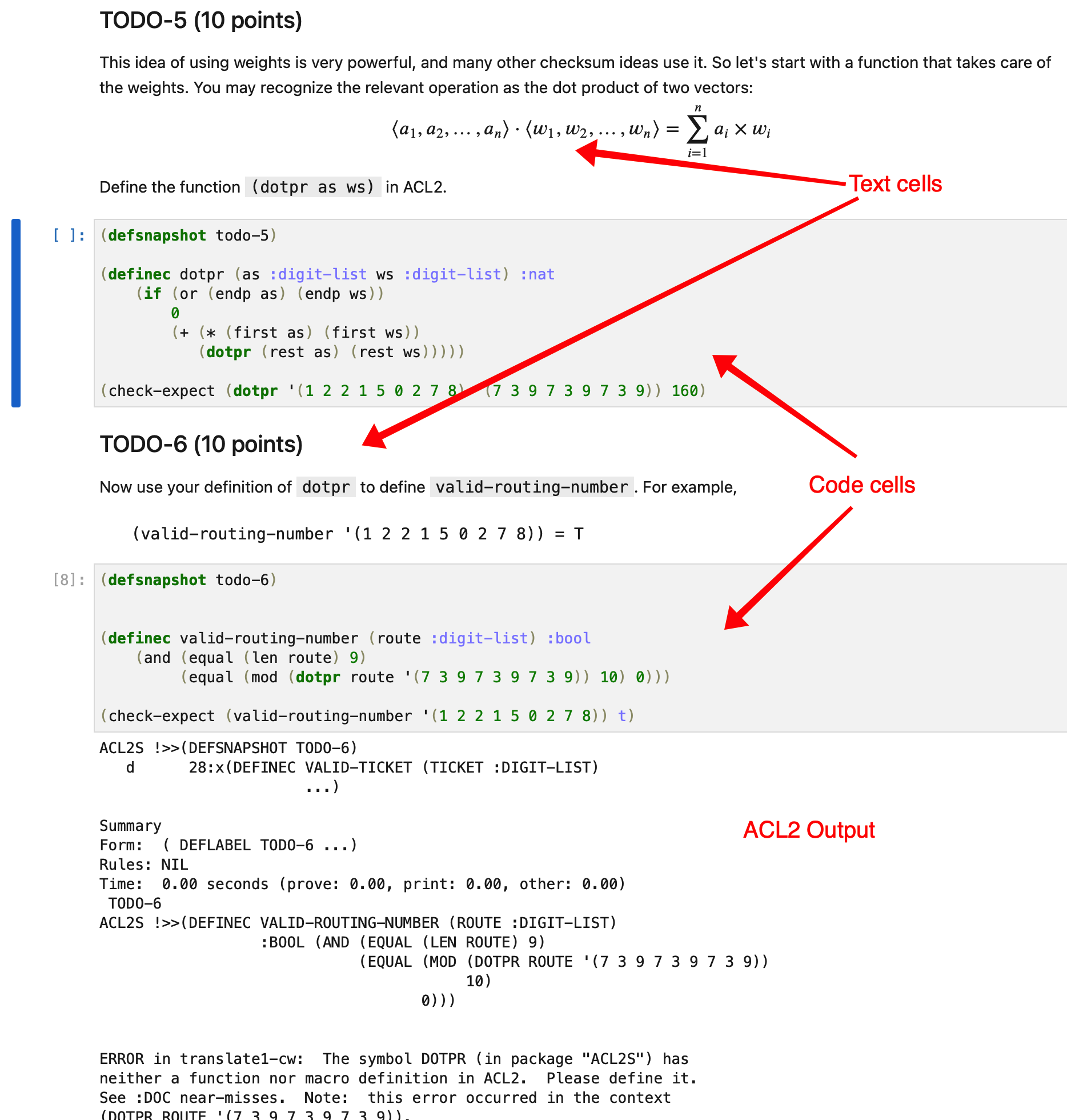}}
\end{center}
\caption{A Jupyter Notebook with ACL2}
\label{fig:jupyter-cell}
\end{figure}

While the Jupyter/Docker solution has been working, it is not without its flaws. One major issue is that students
are sometimes confused by history management between the different cells in a Jupyter notebook. 
Fig.~\ref{fig:jupyter-cell} shows part of the assignment inside a Jupyter notebook. Notice there are blocks of text
(including math) and blocks of code. Each block is called a ``cell'', and code cells can be submitted to ACL2. But
ACL2 sees the code in the order in which it is submitted. In this case, the student submitted
the second code cell and not the first, hence the error that \texttt{dotpr} is not defined.

Ironically, this is
exactly the reason why ACL2s marks part of an ACL2 input file as immutable. But for now, we feel this presents
us with the best solution for the course. One reason is that Jupyter notebooks permit a mixture of explanatory text
and live ACL2 components, which makes them effective for writing both interactive tutorials and assignments. In fact,
that is the way we introduce students to ACL2; instead of lecturing on ACL2 syntax, we give them a suite of 
interactive Jupyter notebooks that slowly walk them through the features they need to learn for this course.

\section{How Much ACL2?}
\label{sec:how-much-acl2}

ACL2 experts may be surprised at how little ACL2 we introduce in this course. Students learn about arithmetic 
built-ins, like \texttt{+}, \texttt{-}, \texttt{*}, \texttt{/}, \texttt{floor}, \texttt{mod}, and \texttt{expt}. 
Students also learn about  \texttt{if}, which we described as similar to C's ternary operator \texttt{( ? : )} 
instead of an ``if statement'', and the comparison operators such as \texttt{=}, \texttt{<}, and friends. The only
other built-ins that students learn about are the list built-in operators, like \texttt{cons}, \texttt{first},
\texttt{rest}\footnote{We avoid \texttt{car} and \texttt{cdr}, though we have to explain them when students see
them in failed proof attempts.}, \texttt{list}, and \texttt{append}. In order to write recursive programs, the
students also have to learn about \texttt{zp} and \texttt{endp}.

To write properties, students learn the basic propositional connectives in ACL2, e.g., \texttt{implies}. All
variables are universally quantified, and students do not learn about explicit quantifiers in ACL2.

Data types are important, and we use mostly numbers, lists of numbers, and arbitrarily nested lists of numbers 
(i.e., trees of numbers). This means that students learn some of the built-in types, e.g., \texttt{:nat},
\texttt{:rational}, \texttt{:nat-list}, and they also learn some of the basics of \texttt{defdata} to construct
new types, such as a list of pairs of naturals, or a list of exactly five naturals.

To define functions, students are taught about \texttt{definec}, instead of \texttt{defun}. The reason, of course,
is that we want students to use types and think about the types of function arguments and return values. That's
one of the lessons we learned from teaching this course multiple times. Students really do find it easier to
program when they know the valid inputs for their functions.

Felleisen observed a similar phenomenon in beginning programmers, and he developed a programming philosophy where
the outline (``recipe'') of the function is determined by its argument's datatypes~\cite{htdp}. E.g.,
oversimplifying a bit, a recipe may say, ``To process a list, do something to the first element of the list, then
recursively  process the rest of the list.'' 

Generally speaking, we are heavily influenced by the work of Felleisen's group on teaching
introductory programming. One aspect that we have shamelessly borrowed from their philosophy is the emphasis on
writing examples before writing code. For instance, we encourage students to begin writing the function 
\texttt{sumlist} by first writing some examples, such as
\begin{lstlisting}
	(sumlist '(1 2 3 4)) = 1 + 2 + 3 + 4
	(sumlist '(  2 3 4)) =     2 + 3 + 4
\end{lstlisting}
The examples by themselves serve an important purpose. All examples can simply become unit test cases. In this
instance, we have identified two test cases, even before writing the code:
\begin{lstlisting}
	(check-expect (sumlist '(1 2 3 4)) (+ 1 2 3 4))
	(check-expect (sumlist '(  2 3 4)) (+   2 3 4))
\end{lstlisting}
Again, this connects programming in ACL2 with industry-approved ``best practices'' such as Test-Driven 
Development~\cite{tdd}.

Examples also have a major secondary purpose, which is that they help students discover the structure of the
code. In fact, our choice of whitespace in those examples was meant to make the outline of the program clear.
Felleisen's programming discipline does this organically, and the notion was reified in the spinoff Bootstrap World,
which users these ideas to teach middle schoolers how to write programs in a dialect of Lisp~\cite{bootstrap}.

We are also influenced by Page's perspective that functional programs are really just equalities, so a definition
should have the form 
\begin{lstlisting}
	(f x) = (... x ...)
\end{lstlisting}
Some readers may recognize this style of expressing definitions as the preferred style in THM and NQTHM~\cite{acl}.
In this case, a suitable defining equation may look like
$$
\texttt{sumlist($L$)} = 
\begin{cases}
0,              & \texttt{if $L=$ NIL} \\ 	
\texttt{first($L$)} + \texttt{sumlist(rest($L$))}, & \text{otherwise}
\end{cases}
$$
Following Felleisen and Page, we insist that students follow a process:
\begin{enumerate}
\item Analyze problem and data definition
\item Determine function contract (i.e., types)	
\item Describe examples
\item Create unit tests (from the examples)
\item Write down definitional equations (inspired by the examples)
\item Write function definition in ACL2
\end{enumerate}
Students are expected to follow this process religiously, and no debugging help is given unless they can show
evidence that they followed the process. I.e., it is only at the end of the process that students may write 
'down actual
ACL2 code, such as the following program:
\begin{lstlisting}
	(definec sumlist (l :nat-list) :nat
	  (if (endp l)
	      0
	      (+ (first l)
	         (sumlist (rest l)))))
\end{lstlisting}
Note that the ACL2 definition is simply the same as the defining equation, but with a different syntax.

Finally, we should mention that students are not at all exposed to The Method\texttrademark, rule classes, 
hints, or
even how to write effective rewrite rules. In fact, most of the time students submit theorems to ACL2 using
\texttt{thm}, so thinking of theorems as rules is irrelevant. They are told that one of the ways that users
guide ACL2 towards a proof is by first proving named theorems (using \texttt{defthm} as opposed to \texttt{thm})
that ACL2 remembers for future use, and in that context they are told just enough rewriting to keep them
from accidentally trying to rewrite \texttt{x} to \texttt{(inv (inv x))}. When we design assignments, we work
very hard to ensure that all theorems students are asked to prove are proved automatically by ACL2---and it helps
that the ACL2 world is almost never polluted by previous theorems the students may have introduced.

\section{An ACL2 Assignment}
\label{sec:acl2-assignment}

In this section, we describe an actual programming assignment from the last iteration of this course. This was
the last assignment that semester, so it showcases the peak of students' competence with ACL2.

As stated in Sect.\ref{sec:intro}, all ACL2 programming assignments in this course have one major goal: connect
formal logic and discrete mathematics to the more practical matter of programming. In addition, most programming
assignments serve a secondary purpose which is to enrich students' experience with specific discrete math topics.
In this case, the featured topic was modular arithmetic, and this assignment was inspired by an assignment from a
since discontinued math course at Luther College~\cite{checksums}.

Modular arithmetic is featured in this assignment as a way to compute checksums to validate numbers. The first
application is to verify airline ticket numbers. These are 15-digit numbers (represented as lists of 15 digits)
where the last digit is computed as
the remainder mod 7 of the first 14 digits read as a single 14-digit number. For example, 
$1 2 3 4 5 \mid 6 7 8 9 0 \mid 1 2 3 4 0$ is a valid airline ticket number because $12,345,678,901,234 \bmod 7 = 0$.

This checksum scheme is designed to correct two typical mistakes that people make when copying numbers by hand:
\begin{enumerate}
\item One digit can be written down incorrectly, e.g, 3 instead of 8.
\item Two adjacent digits can be transposed, e.g, 17 instead of 71.	
\end{enumerate}
Given that, a programmer may make the reasonable assumption that if just one of these mutations occurs, it will be
detected by the checksum. This is the property that can be checked or proved with ACL2.

The first non-trivial challenge that students face is deciding how to model in ACL2 the idea of a mutation. For
instance, one possibility is to write a distance function between ticket numbers that counts the number of mutations
detected form one ticket to the other. The correctness property may then be specified with something like the
following:
\begin{lstlisting}
	(implies (and (valid-ticket ticket)
	              (<= (ticket-distance ticket other-ticket) 1))
	         (valid-ticket other-ticket))
\end{lstlisting}
The problem with this, however, is that the function \texttt{ticket-distance} would be quite complicated,
so ACL2 would have a hard time reasoning about it. Worse, students may define \texttt{ticket-distance} in 
unpredictable ways, which makes it nearly impossible to guarantee that ACL2 will be able to prove any
properties automatically.
Even from the testing perspective there is a major problem here.
How can a testing platform randomly guess a pair of tickets that
differ by one mutation? Most pairs will differ by far more than
that, so the theorem will be true but useless.

We address these concerns by asking students to write one function at a time. In theory, that allows us
to guide them into building a predictable solution\footnote{In practice, students are amazingly creative.}.
In this case, we chose to guide students to build a mutation function that takes in a ticket number and
deliberately introduces one of the errors listed above. So now, the property may be specified as follows:
\begin{lstlisting}
	(implies (valid-ticket ticket)
	         (not (valid-ticket (mutate ticket))))
\end{lstlisting}
One of the most important skills that we want students to gain from this course is the ability to state properties
correctly and effectively. By correctly, we mean that the property really does capture what we are trying
to say. By effectively, we mean that properties should be stated in a way that makes them easier to test or
prove.

At this point, experienced ACL2 users will suspect that there is something wrong with the property above,
and it has to do with the definition of \texttt{mutate}. Since \texttt{mutate} is a \emph{function}, it always
returns the same value for the same input. Since its only input parameter is a ticket number, this implies
that it can only be changed in one way. That's obviously not what we meant---though students are sometimes
confused about this essential nature of functional programming.

What we ask students to do is to construct the mutator function \texttt{(change-nth-digit-to L n d)} which
students define recursively since they do not know about \texttt{update-nth}. So now the property looks like
\begin{lstlisting}
    (implies (valid-ticket ticket)
             (not (valid-ticket (change-nth-digit-to ticket n d))))
\end{lstlisting}
Again, this is not exactly correct. The problem is that \texttt{change-nth-digit-to} is not guaranteed to
actually perform a mutation. It is possible, after all, that the nth element of \texttt{ticket} was already 
\texttt{d}. Students were
warned about this possibility and explicitly asked to address it. Many students did this by checking that
the result of \texttt{change-nth-digit-to} was not equal to the original list L, and some did it by checking
that the nth element of L was not equal to d (the hard way, since they don't know about \texttt{nth} either).
So the property properly stated looks like
\begin{lstlisting}
    (implies (and (valid-ticket ticket)
                  (not (equal (change-nth-digit-to ticket n d) ticket)))
             (not (valid-ticket (change-nth-digit-to ticket n d))))
\end{lstlisting}

So much for stating properties correctly. Unfortunately, this property is not stated effectively, in that it
is not compatible with randomized testing (and completely automated proof is probably out of the question as
well). The problem here is that there are some very important constraints on the possible values of the free
variables \texttt{ticket}, \texttt{n}, and \texttt{d}. Starting with the last two, \texttt{n} is an index into
the list \texttt{ticket}, so since tickets have 15 digits, \texttt{n} must be in the range 0 to 14, inclusive.
And since \texttt{d} is a (wrong) digit, it must be in the range 0 to 9, inclusive. This is the kind of information
that a testing harness needs to know in order to generate reasonable test cases. The counter-example generation
of ACL2s uses predicates in the hypothesis to derive this information, so the property is more effectively
described as follows:
\begin{lstlisting}
    (implies (and (valid-ticket ticket)
                  (natp n)
                  (< n 15)
                  (digitp d)
                  (not (equal (change-nth-digit-to ticket n d) ticket)))
             (not (valid-ticket (change-nth-digit-to ticket n d))))
\end{lstlisting}
Here, \texttt{digitp} is the recognizer for the data type \texttt{:digit}, which students were asked to define
previously. But this still leaves \texttt{ticket}. The function \texttt{valid-ticket} is a regular ACL2 function,
not something known to the counter-example generation framework. The likelihood that a randomly generated ACL2
object will actually be a \texttt{valid-ticket} is essentially zero, so the property as it stands is still not
written effectively. 

A better version would use the fact that all airline tickets are in fact lists of digits. The type 
\texttt{:digit-list} is easy to define using the \texttt{defdata} framework, so the testing framework can
generate random inputs. Recall that the counter-example generation and the data type definition features of ACL2s 
are tightly interconnected. That suggests the following property definition:
\begin{lstlisting}
    (implies (and (digit-listp ticket)
                  (valid-ticket ticket)
                  (natp n)
                  (< n 15)
                  (digitp d)
                  (not (equal (change-nth-digit-to ticket n d) ticket)))
             (not (valid-ticket (change-nth-digit-to ticket n d))))
\end{lstlisting}
When ACL2 is asked to verity this property with random testing, the results are 
not entirely encouraging:
\begin{lstlisting}
	**Summary of Cgen/testing**
	We tested 3000 examples across 3 subgoals, of which 0 (0 unique) 
	satisfied the hypotheses, and found 0 counterexamples and 0 witnesses.

	Test? succeeded. No counterexamples were found.	
\end{lstlisting}
Apparently ACL2 attempted to verify the property 3,000 times and did not find any
counterexamples. The disappointing news, however, is that none of those 3,000 trials
resulted in a viable test. This should not be too surprising. Recall that a ticket
is valid if it has exactly 15 digits and the last digit is the remainder mod 7 of the
first 14 digits taken as a single number. Now, what is the probability that a randomly
generated list of digits is a valid ticket? It's not too surprising that it is somewhat
less that 1/3000, especially considering that a random list is unlikely to have exactly
15 elements.

For numbers, adding a hypothesis such as \texttt{(<= 10000 x)} will ensure that ACL2 only
generates random values that are at least 10,000. But adding the hypothesis 
\texttt{(<= (len l) 15)} does not help for lists. Confronted with this problem, students 
discover that a brute-force approach works. The idea is to define a data type that consists
of lists of precisely 15 digits. Replacing \texttt{(digit-listp ticket)} with
\texttt{(digit-15-listp ticket)} gives more useful results:
\begin{lstlisting}
	**Summary of Cgen/testing**
	We tested 45 examples across 1 subgoals, of which 18 (18 unique) 
	satisfied the hypotheses, and found 3 counterexamples and 15 witnesses.
	
	We falsified the conjecture. Here are counterexamples:
	 [found in : "top"]
	 -- ((D 7) (N 11) (TICKET '(4 2 0 0 0 0 0 0 0 0 0 0 0 0 0)))
	 -- ((D 7) (N 12) (TICKET '(6 4 4 0 7 0 0 0 0 0 0 0 0 0 0)))
	 -- ((D 7) (N 5) (TICKET '(2 8 7 7 0 0 0 0 0 0 0 0 0 0 0)))
	
	Cases in which the conjecture is true include:
	 [found in : "top"]
	 -- ((D 6) (N 4) (TICKET '(0 0 0 0 0 0 0 0 0 0 0 0 0 0 0)))
	 -- ((D 1) (N 6) (TICKET '(0 7 0 7 0 0 0 0 0 0 0 0 0 0 0)))
	 -- ((D 6) (N 3) (TICKET '(7 7 7 7 0 0 0 0 0 0 0 0 0 0 0)))
	
	Test? found a counterexample.
\end{lstlisting}
A few points are worth mentioning. First, ACL2 only generated 45 test cases for this,
instead of 3,000 for the previous version. This seems to be related to the data type
of lists with precisely 15 elements. Experimenting with lists of different, but still
fixed length suggests that the longer the lists, the fewer test cases are generated.
Second, out of those 45 test cases 18 satisfied the hypotheses. Since the only hypothesis
that is not already satisfied by the test case generation is the actual checksum, we would
have expected that only 1 out of 7 tests should have actually satisfied the hypothesis,
as opposed to more than 1 in 3. Precisely why so many random tickets are actually valid is
a mystery. Finally, looking at the generated ticket numbers, it is
obvious that ACL2 is favoring ticket numbers that end in a string of zeros. 

Those would be important issues for professional programmers. The big lesson for students
here is that they must always learn the capabilities and limitations of their tools, and
that part of being a professional is to understand how to exploit the capabilities and
manage the limitations. Since proficiency in ACL2 and the ACL2 toolchain is not one of the
student learning outcomes for this class, we do not dwell on these issues here. Rather, we
want students to learn about the value of testing in whatever platform they use.

This particular example shows just how useful randomized testing can be,
even with just a limited number of test cases. 
ACL2 found three counter examples, which is enough to invalidate our
expectation. We know now that there are some 1-digit errors that are not detected by this
checksum scheme. Looking at these three counter-examples, it is easy to see a pattern.
In each case, a 0 was replaced by a 7. That is enough to suggest what the real problem is:
This checksum scheme can not detect an error in which a digit is replaced by another digit 
that is equivalent mod 7, and this suspicion can be confirmed by thinking about modular
arithmetic. Randomized testing proved its worth by detecting errors in our assumptions and
also suggesting how those assumptions can be revised. 

The rest of the assignment considers different algorithms for computing checksums. 
One of these is the checksum for bank routing numbers, which are 9-digit numbers
such that 
$7a_1 + 3a_2 + 9a_3 + 7a_4 + 3a_5 + 9a_6 + 7a_7 + 3a_8 + 9a_9 \equiv 0 \pmod {10}$, 
where $a_i$ is the i\textsuperscript{th} digit. This algorithm seems conceptually
different, in that they use weights to multiply the digits, as opposed to reading
the digits as a single number. This distinction evaporates when you consider that
a multi-digit number is really the weighted sum of its digits, where the weights
are powers of 10.

The code that computes checksums is relatively trivial, and the properties that state
that checksums can detect a 1-digit change are identical to the case with airline
ticket numbers, so we will not discuss either issue in this paper. However, we will
mention that ACL2 tries to validate this property with 12,000 random 
inputs\footnote{There is something we simply don't understand about how the 
counter-example generation algorithm works. E.g., why does ACL2s pick 12,000 examples
here, but only 45 for airline ticket numbers?}, out of which 948 satisfied the
hypotheses. This is close to the 1/10 ratio we would expect for a checksum based on
mod 10, and the difference is partially explained by the fact that sometimes (10\% of
the time), a mutation does not result in a different routing number. The most important 
bit of information, however, is that \emph{all} of those
1-digit errors were detected by the checksum algorithm, so we expect that this property
is in fact true. Readers familiar with basic number theory will correctly surmise that
this is the case because the weights (7, 3, and 9) are all relatively prime to the
modulus 10.

That goes for mutating a routing number by changing one of its digits. However, the
picture is different when we consider transposing two digits. ACL2 verifies this
property by testing 1,662 input values, only 32 of which satisfy the hypothesis. It is
worth considering why so few items satisfy the hypothesis, especially since this is
so much different than in the previous case. We believe the reason is that ACL2 is
selecting so many tickets with long sequences of zeros. Replacing a 0 by an adjacent 0
does not change the routing number, so the rate of distinct mutations is much lower than
the expected 1/10. But out of those 32 actual test cases, ACL2 is able to to detect three
cases in which the mutation was undetected by the checksum. Once again, randomized testing 
proves its worth:
\begin{lstlisting}
	**Summary of Cgen/testing**
	We tested 1662 examples across 2 subgoals, of which 32 (32 unique)
	satisfied the hypotheses, and found 3 counterexamples and 29 witnesses.
	
	We falsified the conjecture. Here are counterexamples:
	 [found in : "top"]
	 -- ((N 0) (ROUTE '(3 8 8 0 0 7 0 0 0)))
	 -- ((N 0) (ROUTE '(6 1 4 7 0 0 0 0 0)))
	 [found in : "Goal"]
	(IMPLIES (AND (ROUTING-NUMBERP ROUTE)
	              (VALID-ROUTING-NUMBER ROUTE)
	              (NATP N)
	              (< N 8)
	              (NOT (EQUAL (TRANSPOSE-NTH-DIGIT ROUTE N)
	                          ROUTE)))
	         (NOT (VALID-ROUTING-NUMBER (TRANSPOSE-NTH-DIGIT ROUTE N))))
	
	 -- ((N 1) (ROUTE '(4 9 4 7 0 0 0 0 0)))
	
	Cases in which the conjecture is true include:
	 [found in : "top"]
	 -- ((N 1) (ROUTE '(3 1 5 0 7 0 0 0 0)))
	 -- ((N 4) (ROUTE '(4 1 1 7 7 0 0 0 0)))
	 -- ((N 1) (ROUTE '(7 7 0 0 0 0 0 0 0)))
	
	Test? found a counterexample.
\end{lstlisting}
Looking at the counterexamples, again a pattern emerges. The undetected errors
occurred when the digits swapped were 3 and 8, 6 and 1, or 4 and 9. Once more,
readers with a background in number theory may recognize that in all of these
cases the difference between the digits is a multiple of 5, and 5 divides the
modulus 10.

Students are asked to consider other checksum schemes. The scheme used to validate credit
card numbers is very interesting, because it manages to disguise the use of modular
arithmetic by introducing a curious counting scheme that deals effectively with the 
case that a digit is replaced with another digit from which it differs by 5. The
scheme also accounts for most transposition errors, with the only undetected changes
happening when a 0 and 9 are transposed. The important thing here, from our
perspective, is that ACL2 confirms that all 1-digit substitution errors are detected,
and sometimes (but not always) it manages to find counter-examples for a single 
transposition. Again, the value of randomized testing is clear, though students 
learn not to become complacent, since some errors occur rarely enough that they can
slip through randomized testing.

Another scheme that students examine is the one used in ISBN-10 codes. This one uses
mod 11 arithmetic, not mod 10. One disadvantage of this is that the checksum is a digit
from 0 to 10, inclusive, which explains why many books have a trailing ``X'' in their
ISBN. But the big advantage of this scheme is that since 11 is a prime number,
it is relatively prime to any weight used on the digits, and it is also relatively
prime to the difference of adjacent weights. So this scheme is able to detect all single
errors in transmission, whether due to a single digit change or a transposition.
As expected, ACL2 does not find any counterexamples when this checksum scheme is used.

What students get from this assignment is a deeper appreciation for software testing and
the benefits of incorporating testing into the development process. Students also learn
to explore and understand the capabilities of their tools, so they can predict what types
of errors are still likely to get through. In other assignments, students see how 
systems like ACL2 can be used to prove correctness of programs, gaining the highest
possible level of assurance in the software. However, proofs have not been featured in
this assignment, for two main reasons. The first, of course, is that so many of the 
properties
are actually false. The various checksum schemes have holes through which single errors
in transmission can get through undetected. But even when the checksum scheme is strong
enough to detect all single errors, the proof requires reasoning about basic facts from
number theory. This is well within the capabilities of ACL2, but not on its own. These
proofs require the guidance of a user with significant experience in ACL2 formalizations,
and that is beyond the scope of this class.

Readers experienced with ACL2  may wonder whether any properties are ever simple enough
that ACL2 can prove them on their own. To address that concern, we describe the last
part of this assignment. The US Postal Service uses a barcode to encode some address
information, such as the street number and the zipcode. This barcode uses long and short
bars to represent digital information. Each digit is encoded into five bits, which allows
for significant redundancy. The encoding for each bit consists of three 1s and two 0s,
so a single-bit error can be detected at the digit level. Moreover, a checksum is added
to the list of digits. If a single bit is flipped, it is possible (1) to determine which
digit was corrupted, and hence (2) to correct that digit by recomputing the checksum.

The first step to implement this scheme in ACL2 is to define the function 
\texttt{encode} that takes a list of digits and converts it into a list of
bits, and the inverse function \texttt{decode}. Although there are fancier and
better ways of writing these functions, students are encouraged to do this using 
nested IFs. Then they are asked to prove using ACL2 that \texttt{encode} and
\texttt{decode} are, in fact, inverses of each other. The proof requires an 
induction over the input list, and the key step at the induction explores the
nested IFs. There are many cases to consider, but each of the cases is trivial.
ACL2 proves this completely automatically, demonstrating resilience even among
the variety of definitions that students tend to write.

\section{Conclusions}
\label{sec:conclusions}

This paper described our experience using ACL2 in a course on Discrete Mathematics for
(mostly) computer science majors. The course is focused on Discrete Math topics, not ACL2;
we spend only one out of 15 weeks of lecture in ACL2 itself. But we use ACL2 to drive a
few points:
\begin{itemize}
\item Logic is a tool that can be useful even to the most pragmatic professional 
		programmer.	
\item Proof is a technique that can be used to guarantee software correctness, and it sits
		on a spectrum of techniques that can raise our confidence that our software works.
		Some of these techniques, like unit testing and randomized testing, are already
		widely accepted in industry.
\item Although reasoning about programs at scale is very difficult, it is possible to 
		mechanize much of logic and build tools that can assist in formal reasoning.
		ACL2 is one of these tools, and it is currently used in industrial applications.
\end{itemize}
We believe that these three principles are uncontroversial among those in our
community. At the very least, no students are harmed by light exposure to ACL2 and formal
methods, and in a different context we even convinced our university's Institutional 
Review Board of this simple fact. We do not expect our students to find jobs that require
formal methods upon graduation, but at the very least we do expect them to incorporate
serious and extensive testing practices as they mature into \textbf{professional} software
developers. And we hope that in five to ten years' time, when they assume leadership
positions, they will remember that formal methods can help to develop software that meets
the highest standards of reliability.

\nocite{*}
\bibliographystyle{eptcs}
\bibliography{teaching-testing}
\end{document}